\documentclass{aa}  
\usepackage{graphicx}  
\newcommand{\ltsima} {$\; \buildrel < \over \sim \;$}  
\newcommand{\gtsima} {$\; \buildrel > \over \sim \;$}  
\newcommand{\lta} {\lower.5ex\hbox{\ltsima}}  
\newcommand{\gta} {\lower.5ex\hbox{\gtsima}}  
\begin{document}  
\title{Understanding the nature of FR~II optical nuclei: a new diagnostic
plane for radio galaxies\thanks{Based  on observations obtained at
the  Space  Telescope Science  Institute,  which  is  operated by  the
Association of  Universities for Research  in Astronomy, Incorporated,
under NASA contract NAS 5-26555.}} 
  
\titlerunning{The nuclei of FR~II radio galaxies}  
  
  
\author{M. Chiaberge   
\inst{1,2}\fnmsep\thanks{ESA fellow}
\and    
A. Capetti \inst{3}    
\and   
A. Celotti \inst{4}}   
   
\offprints{M. Chiaberge}

\institute{Space Telescope  Science Institute, 3700  San Martin Drive,
Baltimore, MD  21218, USA \\ \email{chiab@stsci.edu}  \and Istituto di
Radioastronomia del  CNR, Via P.  Gobetti 101, I-40129  Bologna, Italy
\and  Osservatorio  Astronomico  di  Torino, Strada  Osservatorio  20,
I-10025  Pino Torinese,  Italy  \\ \and  SISSA/ISAS,  Via Beirut  2-4,
I-34014 Trieste, Italy \\}
  
\date{Received 30 November 2001; accepted 28 June 2002}  
   
\abstract{We  extend  our study  of  the  nuclei  of 3CR  FR~II  radio
galaxies through HST optical images up  to $z = 0.3$.  In the majority
of them  an unresolved nucleus  (central compact core, CCC)  is found.
We analyze their position in the plane formed by the radio and optical
nuclear luminosities in relation to their optical spectral properties.
The  broad--lined objects (BLO)  have the  brightest nuclei:  they are
present  only  at  optical  luminosities  $\nu  L_{\nu}  \gta  4\times
10^{42}$ erg s$^{-1}$ which we  suggest might represent a threshold in
the radiative  efficiency combined  with a small  range of  black hole
masses.  About  $40 \%$ of the  high and low  excitation galaxies (HEG
and LEG)  show CCC  which resemble those  previously detected  in FR~I
galaxies,  in  apparent  contrast   to  the  unification  model.   The
equivalent  width of  the [OIII]  emission line  (with respect  to the
nuclear  luminosity) reveals  the nature  of these  nuclei, indicating
that the  nuclei of  HEG are obscured  to our  line of sight  and only
scattered radiation is observed.   This implies that the population of
FR~II is  composed of objects  with different nuclear  properties, and
only   a   fraction   of   them   can   be   unified   with   quasars.
}
  
   \maketitle  
%

\section{Introduction}  
  
In  the  framework  of  the  AGN unification  scheme  for  radio--loud
sources,   powerful  radio   galaxies   with  FR~II   edge--brightened
morphology (Fanaroff \& Riley \cite{fr}) are believed to be misaligned
quasars, while lower power, edge--darkened FR~I are associated with BL
Lac objects.  This  basic unification picture (Barthel \cite{barthel},
Orr \& Browne \cite{browne}, see Urry \& Padovani \cite{urrypad} for a
review)  is  mainly  supported  by  the  comparison  of  the  extended
properties (radio  morphology and linear dimensions,  host galaxy type
and environment,  and narrow emission  line luminosity) as well  as by
the  number counts  of the  two  classes.  However,  the detection  of
polarized broad  emission lines, although  only for a small  number of
FR~II, is the  most direct evidence of the  unification scheme and the
presence  of  absorbing ``tori''  in  high  power  radio loud  sources
(Antonucci   \&    Barvainis   \cite{antonucci90},   Cohen    et   al.
\cite{cohen99}).
 
How these  results extend  to low power  sources is still  unclear. In
fact, on large scales,  the morphological FR~I/FR~II dicothomy appears
to be also  associated with other (large scale)  properties.  From the
optical point of view, FR~II are associated with different sub-classes
of  bright elliptical  galaxies.   On average,  FR~II  hosts are  less
luminous with respect  to FR~I ones (Owen \cite{owen},  Ledlow \& Owen
\cite{ledlowowen}), and  belong to lower  density groups, at  least at
low redshifts (e.g.  Zirbel \cite{zirb96}, \cite{zirb97}).

However,  how  these large--scale  properties  relate  to the  nuclear
structure  and  central  activity   is  still  a  debated  issue.   In
particular,  the properties  of  the emission  lines  observed in  the
spectrum  of radio  galaxies and  plausibly connected  to  the nuclear
activity, have revealed a phenomenology richer than the radio one.
For example,  the role  of a sub-class  of low-ionization  FR~II (e.g.
Laing et al.  \cite{laing94}) has still to be assessed.  These objects
have an  FR~II morphology, but  their optical spectral  properties are
similar to those of  FR~I.  Wall \& Jackson (\cite{jacksonwall97}) and
Jackson  \&  Wall (\cite{jacksonwall99})  proposed  that such  objects
constitute, together with FR~I, a single population of radio galaxies.
Furthermore, Willott et al.  (\cite{willott}) have recently found that
the fraction of  objects with observed broad lines (in  the 6C, 7C and
3CRR  samples and  having  excluded FR~I)  decreases with  luminosity.
They indicate as  a possible explanation for this  lack of quasars the
rise of  a distinct population of  radio galaxies which  have an FR~II
radio morphology but lack a well-fed quasar nucleus.
  
Our  aim is to  investigate these  issues by  directly looking  at the
nuclear  continuum  emission  in  the optical  band,  identifying  its
physical origin  and relating it to  both the radio  and emission line
properties.   HST  optical images  of  the  nuclear  regions of  radio
galaxies are suited to this goal as their high resolution allows us to
separate the AGN emission from the stellar host galaxy background.  In
particular, the  optical snapshot surveys of 3CR  objects (e.g. Martel
et al.  \cite{martel},  De Koff et al.  \cite{dekoff})  has provided a
wealth of high quality data for this purpose.
  
We have started this study by considering complete samples of FR~I and 
FR~II from this catalog (Chiaberge et al. \cite{pap1}, hereafter 
Paper~I; Chiaberge et al.  \cite{pap2}, hereafter Paper~II).  We 
analyzed the properties of unresolved optical nuclei, which have been 
found to be present in the great majority of the objects. 
  
The  optical  nuclei  of  3CR  FR~I behave  similarly,  and  are  best
explained as  non thermal  synchrotron emission from  the base  of the
relativistic jet.  Furthermore, the  high detection rate ($\sim 85\%$)
directly  implies  that geometrically  thick  obscuring  tori are  not
present in  FR~I radio galaxies  (or, alternatively, they  are present
only in a  minority of them).  Given this, the  lack of broad emission
lines in these objects cannot be due to obscuration.
  
The behavior of FR~II (at redshift below $z=0.1$) appears to be more 
complex, although their properties are clearly related to their 
spectral classification. Broad line radio galaxies show an optical 
excess with respect to the expected non thermal emission level, which 
might indicate a contribution from the thermal disk.  Several radio 
galaxies in which broad lines are absent do not show any nuclear 
source, and they can be interpreted as obscured nuclei, as expected in 
the framework of the current AGN unification scheme.  Most importantly, 5 
sources of the sample have a core with radio-optical properties that 
are completely consistent with those found in FR~I.  Although not all 
of them belong to the low--ionization subclass, it is tempting to 
consider them as FR~I--like. 
  
In all cases, this nuclear emission sets an upper limit to any 
radiation from the accretion flow. In particular, in the non--thermally 
dominated nuclei (FR~I or FR~II) this seems to imply that 
accretion might take place on a low efficiency radiative regime. 
  
Since it is important both to establish whether these findings hold 
only for nearby sources, and to analyze a larger sample of objects in 
order to improve the statistics, in this paper we extend the sample up 
to a redshift of $z=0.3$. Furthermore we will show that in order to 
address the nature of the nuclei, a crucial parameter is the 
equivalent width of the [OIII] emission line, which we calculate with 
respect to the nuclear continuum emission.  The organization of the 
paper is as follows: in Sect.  \ref{sample} we describe our sample of 
FR~II and the HST observations; in Sect.  \ref{fr2ccc} we present the 
results of the photometry of the nuclei and we analyze the relation 
between the optical and radio core luminosity; in Sect. 
\ref{discussion} we discuss our results for the different spectral 
subclasses, also considering their radio properties and, most 
importantly, the [OIII] emission line luminosity.  In Sect. 
\ref{conclusions} we present a summary of our findings and we draw 
conclusions and future perspectives. 
  
$H_0= 75$ km s$^{-1}$  Mpc$^{-1}$ and $q_0=0.5$ are adopted throughout
the paper.  The  spectral index $\alpha$ is defined  as $F_\nu \propto
\nu^{-\alpha}$.
  
  
\section{The sample and the HST observations}  
\label{sample}  
  
The sample considered here comprises all radio galaxies belonging to  
the 3CR catalogue (Spinrad \cite{spinrad}) with redshift $z<0.3$,  
morphologically classified as FR~II.  We directly checked their  
classification for erroneous or doubtful identifications by searching  
the literature for the most recent radio maps.  The final list  
constitutes a complete, flux and redshift limited sample of 65 FR~II  
radio galaxies.   
  
We  searched  for   optical  spectral  classification  and/or  optical  
spectra, in  order to  differentiate our sources  on the basis  of the  
presence of broad or narrow and high or low excitation emission lines.  
Here   we  adopt   the  scheme   defined  by   Jackson   and  Rawlings  
(\cite{jackson}). They classify as  WQ (weak quasar) objects for which  
at least one broad line  has been observed and whose optical continuum  
V-band luminosity  is $<10^{23}$ W Hz$^{-1}$ while  quasars (QSO) have  
brighter continua; high and  low ionization narrow-lined galaxies (HEG  
and LEG) are classified on  the basis of their [OIII] equivalent width  
(less than  10 \AA ~for a  LEG) and/or [OII]/[OIII] ratio  ($>1$ for a  
LEG).  In the  following, we  will also  refer to  WQ and  QSO  as the  
broad-lined objects (BLO).  
  
In Table  \ref{tab1} we show  redshifts and radio data  of objects  
with $z<0.3$, as taken from  the literature, together with the optical  
spectral  classifications. Low  redshift objects  ($z<0.1$)  have been  
already discussed in Paper~II and are listed here for completeness.

\begin{figure}  
   \resizebox{\hsize}{!}  
   {\includegraphics[width=\textwidth]{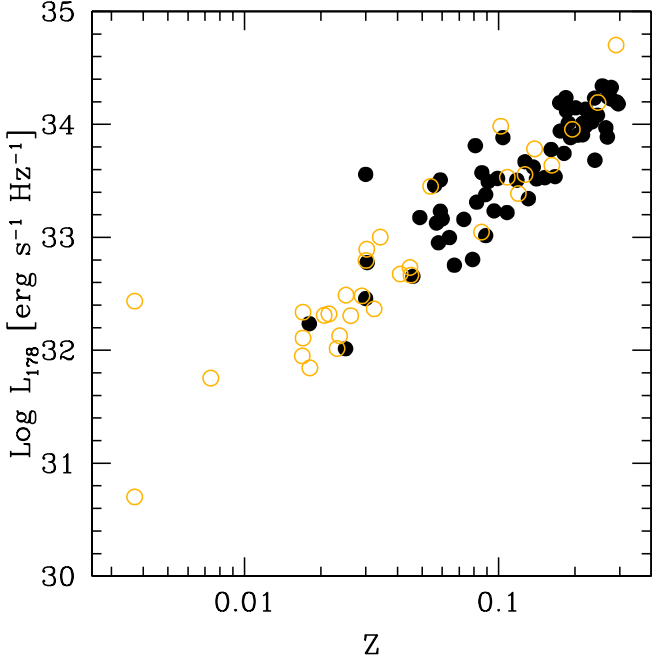}}  
   \caption{Total radio luminosity at 178 MHz vs. redshift. FR~II are  
plotted as filled circles, while FR~I are empty circles.}  
              \label{lumz}  
    \end{figure}

In  Fig. \ref{lumz}  we show  the redshift  vs total  radio luminosity  
diagram for the sample of  FR~II galaxies, together with the sample of  
FR~I discussed in Paper~I. FR~II have a median redshift $z=0.152$, and  
total  radio  luminosities  at  178  MHz  are  between  $10^{32}$  and  
10$^{34.4}$ erg  s$^{-1}$ Hz$^{-1}$.  Note  that  whereas the  two  
samples are  selected at the same  limits of redshift  and flux, FR~II  
are,  on average, more  luminous and  distant than  FR~I. Furthermore,  
there is  no significant difference  in the distributions  of extended  
radio power for the different spectral classes of FR~II.

\begin{table*}
\caption{The sample of FR~II radiogalaxies}
\tiny
\begin{tabular}{l c c c c c| l c c c c c} \hline

Source Name    &  redshift & Class. & $S_t$ (178)   &  $\log L_r$             &  $\log L_{[OIII]}$ & Source Name    &  redshift & Class. & $S_t$ (178)   &  $\log L_r$             &  $\log L_{[OIII]}$ \\      
               &   $z$     &        &     Jy       & erg s$^{-1}$ Hz$^{-1}$  &   erg s$^{-1}$     &                &   $z$     &        &      Jy       & erg s$^{-1}$ Hz$^{-1}$  &   erg s$^{-1}$     \\     
               &           &        &               &              &                    &                       &           &        &               &              &       \\
   3C~15       &   0.073   &   LEG  &      15.8     &   31.54      &       40.38        &           3C~227$^a$  &   0.086   &   WQ   &      30.0     &   30.48      &       41.78          \\  
   3C~17       &   0.220   &   WQ   &	   20.0     &   32.76	   &	   41.91    	&   	    3C~234      &   0.185   &   WQ   &      26.1     &   31.88      &       43.17          \\  
   3C~18       &   0.188   &   HEG  &	   19.8     &   31.84	   &	   41.88    	&  	    3C~236      &   0.099   &   LEG  &      20.5     &   31.51      &       40.47          \\  
   3C~33       &   0.059   &   HEG  &	   53.0     &   30.27	   &	   42.01    	&   	    3C~284      &   0.239   &   HEG  &       6.1     &   30.25      &       42.19          \\  
   3C~33.1     &   0.181   &   WQ   &	   11.4     &   31.03	   &	   41.66    	&  	    3C~285      &   0.079   &   HEG  &      6.0      &   29.93      &       40.73          \\  
   3C~35       &   0.067   &   LEG  &	    7.3     &   30.27	   &	   40.03    	&  	    3C~287.1    &   0.216   &   QSO  &      14.2     &   32.53      &        --            \\  
   3C~40       &   0.018   &   LEG  &	   28.6     &   30.60	   &	   38.85    	&  	    3C~300      &   0.270   &   HEG  &      17.97    &   31.07      &       42.16          \\  
   3C~61.1     &   0.184   &   LEG  &	   34.51    &   30.33	   &	   42.09    	&  	    3C~303      &   0.141   &   WQ   &      10.6     &   31.80      &        --            \\  
   3C~63       &   0.175   &   HEG  &	   19.1     &   30.97	   &	    --      	&  	    3C~318.1    &   0.046   &   LEG  &      12.0     &    --        &        --            \\  
   3C~79       &   0.256   &   HEG  &	   24.8     &   31.19	   &	   42.43    	&  	    3C~319      &   0.192   &   LEG  &      14.2     &   29.93      &        --            \\  
   3C~88       &   0.030   &   LEG  &	   17.5     &   30.50	   &	   39.78    	&  	    3C~321      &   0.096   &   HEG  &      11.2     &   30.78      &       42.15          \\  
   3C~98       &   0.030   &   HEG  &	   35.5     &   29.25	   &	   40.87    	&  	    3C~323.1    &   0.264   &   QSO  &      10.0     &   31.69      &       42.93          \\  
   3C~105      &   0.089   &   HEG  &	   18.0     &   30.36	   &	   40.78    	&  	    3C~326      &   0.089   &   LEG  &       7.8     &   30.34      &       40.64          \\  
   3C~111      &   0.049   &   WQ   &	   35.2     &   31.69	   &	    --      	&  	    3C~327      &   0.104   &   HEG  &      43.0     &   30.88      &       42.05          \\  
   3C~123      &   0.218   &   LEG  &	  189.0     &   31.82	   &	   41.58     	&  	    3C~332      &   0.152   &   QSO  &       9.5     &   30.65      &        --            \\  
   3C~132      &   0.214   &   LEG  &	   12.4     &   31.40	   &	    --      	&  	    3C~349      &   0.205   &   HEG  &      13.3     &   31.18      &       41.42          \\  
   3C~133      &   0.277   &   --   &	   21.0     &   32.32	   &	    --      	&  	    3C~353      &   0.030   &   LEG  &      220.0    &   30.54      &       39.34          \\  
   3C~135      &   0.127   &   HEG  &	   18.2     &   30.18	   &	    --      	&  	    3C~357      &   0.167   &   HEG  &       8.2     &   30.48      &        --            \\  
   3C~136.1    &   0.064   &   --   &	   14.0     &    --	   &	    --      	&  	    3C~379.1    &   0.256   &   HEG  &       --      &    --        &        --            \\  
   3C~153      &   0.277   &   HEG  &	   16.67    &   29.73	   &	    --      	&  	    3C~381      &   0.161   &   HEG  &      15.2     &   30.48      &       42.37          \\  
   3C~165      &   0.296   &   LEG  &	   13.5     &   31.08	   &	    --      	&  	    3C~382      &   0.058   &   WQ   &      15.2     &   31.13      &       41.52          \\  
   3C~166      &   0.245   &   --   &	   14.7     &   32.73	   &	    --      	&  	    3C~388$^b$  &   0.091   &   LEG  &      22.5     &   31.04      &       40.49          \\  
   3C~171      &   0.238   &   HEG  &	   21.8     &   30.36	   &	   42.50    	&  	    3C~390.3$^b$ &   0.056   &   WQ   &      52.4     &   31.38      &       42.05          \\  
   3C~173.1    &   0.292   &   LEG  &	   14.3     &   31.18	   &	   41.35    	&  	    3C~401      &   0.201   &   LEG  &      24.0     &   31.50      &       41.01          \\  
   3C~180      &   0.220   &   --   &	   14.2     &    --   	   &	    --    	&  	    3C~402      &   0.025   &   HEG  &      9.0      &   29.73      &        --            \\  
   3C~184.1    &   0.118   &   HEG  &	   14.3     &   30.25	   &	   42.19    	&  	    3C~403      &   0.059   &   HEG  &      28.0     &   29.87      &       41.55          \\  
   3C~192      &   0.060   &   HEG  &	   23.2     &   29.73	   &	   41.60    	&  	    3C~430      &   0.056   &   --   &       --      &    --        &        --            \\  
   3C~197.1    &   0.131   &   HEG  &	    8.1     &   30.30	   &	    --      	&  	    3C~436      &   0.215   &   HEG  &      15.7     &   31.21      &       41.52          \\  
   3C~198      &   0.082   &   HEG  &	   18.0     &    --	   &	   41.04    	&  	    3C~445$^{a,b}$  &   0.057   &   QSO  &      23.5     &   31.34      &       42.05          \\  
   3C~219      &   0.174   &   WQ   &	   34.3     &   31.54	   &	   41.62    	&  	    3C~452      &   0.081   &   HEG  &      58.3     &   31.24      &        --            \\  
   3C~223      &   0.137   &   HEG  &	   14.2     &   30.57	   &	   42.17    	&  	    3C~456      &   0.233   &   HEG  &      14.0     &   31.39      &       42.59          \\  
   3C~223.1    &   0.108   &   HEG  &	   8.7      &   30.24	   &	   41.65    	&  	    3C~458      &   0.289   &   --   &       --      &    --        &        --            \\  
		&          &        &               &              &                    &	    3C~460      &   0.268   &   HEG  &       8.1     &   31.39      &       41.67          \\

\hline                                                                        				  
                                                                              				  
\end{tabular}                                                                 
\label{tab1}

\medskip

Redshifts and  $S_t$, the total radio  flux at 178 MHz,  are from NED;
optical   spectral  classification   is  from   Jackson   \&  Rawlings
\cite{jackson}       and      C.        Willott's       web      page,
http://www-astro.physics.ox.ac.uk/~cjw/3crr/3crr.html;  $L_r$  is  the
radio core  luminosity at 5  GHz (data are  taken from Zirbel  \& Baum
\cite{zirbel95}),    except   for    3C~123    (Hardcastle   et    al.
\cite{hardcastle98}, 8.44 GHz),  3C~133 (Nilsson \cite{nils98}, 5GHz),
3C~153  (Hardcastle  et al.   \cite{hardcastle98},  8.44 GHz),  3C~165
(Nilsson   \cite{nils98},  5   GHz),  3C~223.1   (Hardcastle   et  al.
\cite{hardcastle98},   8.35   GHz);    3C~349   (Hardcastle   et   al.
\cite{hardcastle98},  8.44  GHz).   Radio  core  data  at  frequencies
different  from  5   GHz  were  converted  to  5   GHz  using  a  flat
($\alpha_r=0$) spectral index.  $L_{[OIII]}$  is the luminosity of the
[OIII]  line  (again  from  Jackson  \&  Rawlings  \cite{jackson}  and
Willott's  web page,  except  for  3C~88, which  has  been taken  from
Tadhunter et al. \cite{tadhunter93}).  (a) line subtracted (see text);
(b) saturated core (see Paper~II for  an outline of the method used to
estimate the flux).

\end{table*}

HST observations were performed during the 3CR snapshot survey (Martel  
et al., etc etc.)  and are  available in the public archive for all 65  
objects except  4 (3C~33,  3C~61.1 and 3C~105,  3C~458 which  were not  
observed). The whole sample was observed using the F702W filter  
except for 3C~192 (F555W). Exposure  times are in the range 140-300 s.  
The  data  were  processed  through  the  standard  PODPS  (Post  
Observation Data Processing System) pipeline for bias removal and flat  
fielding (Biretta et al. \cite{biretta}). Individual exposures in each  
filter were combined to remove cosmic rays events.  
  
\section{Optical cores in FR~II and their relationship with radio cores.}  
\label{fr2ccc}  
  
We derive the  radial brightness profiles of all  objects, looking for
unresolved nuclear  components (central compact cores,  CCC).  Here we
adopt the  same definition  for CCC  as in Paper~I  and II  (a central
component with measured FWHM $< 0.08^{\prime\prime}$, i.e.  unresolved
at the  HST resolution). In the $z<0.3$  sample we find 34  CCC and 19
objects  with   smooth  nuclear  profiles  and   no  detected  central
source. In these  objects, we evaluate the upper  limit to any central
component measuring  the light excess  of the central 3x3  pixels with
respect  to the  surrounding galaxy  background.  In  8  galaxies, the
presence  of  complex  morphologies  prevents  us  from  evaluating  a
reliable upper limit to any central source.
  
Considering each spectral class of  FR~II separately, we find that all  
objects classified as  QSO and WQ have a  CCC; among the narrow--lined  
objects, 15 HEG (and 4 LEG) have a  CCC, for 10 HEG (and 9 LEG) we are  
able to set upper limits to  the optical nuclear emission, while 3 HEG  
(and  2 LEG) have  complex nuclear  morphologies. Of  the unclassified  
sources, 2 have a CCC and 3 have complex nuclei.

\begin{table}  
\caption{CCC in FR~II radio galaxies}  
\begin{tabular}{l c c c c c} \hline  
  
Class.           &   CCC  &  Upper lim.  &  Complex  & Not obs. &  Tot. \\  
                 &        &                &           &              &       \\  
QSO              &    4   &       0        &     0     &    0         &  4    \\  
WQ               &    9   &       0        &     0     &    0         &  9    \\  
HEG              &   15   &      10        &     3     &    2         &  30   \\  
LEG              &    4   &       9        &     2     &    1         &  16   \\  
Unclass.         &    2   &       0        &     3     &    1         &  6    \\   
  
\hline  

\end{tabular}                                                                   
\label{tabccc}  
\end{table}

We  performed  aperture photometry   of the  34  CCC,  using the  same
technique as described in Paper~I.  We evaluate the background level at
a distance of  $\sim 5$ pixels  ($\sim 0.23^{\prime\prime}$)  from the
center.   The dominant photometric error  is thus the determination of
the background  in regions of  steep brightness  gradients, especially
for the faintest cores, resulting in a typical error of $\sim 10\%$.
  
All  images were taken with  broad band filters  which, within our
redshift   range,  include  emission   lines  (H$\alpha$   +  [N~II]).
Unfortunately, no HST narrow band images are available for our sources
in  this spectral  range. Since  the broad  line emission  is probably
cospatial to the optical CCC,  we corrected the broad band fluxes from
line contamination in the case of two objects (3C~227 and 3C~445), for
which ground  based measurements of  the broad H$\beta$ line  flux are
available  (Tadhunter  et  al.   \cite{tadhunter}), assuming  a  ratio
H$\alpha$/H$\beta=6$  (Netzer  \cite{netzer}).   In these  cases,  the
resulting emission  line contribution is  $15-20$\% of the  total flux
measured in  the F702W filter\footnote{Other authors  (e.g. Heckman et
al.   \cite{heckman89}) adopt a  value of  H$\alpha$/H$\beta\sim2$ for
radio  galaxies.  Therefore  our estimate  might be  considered  as an
upper limit  to the  line contamination.}. We  have not  corrected the
flux of  the other QSO  and WQ, however,  we have tested that  the the
total $H\alpha$ line flux (Baum \& Heckman \cite{baumheckman89}) would
represent only $10-20\%$ of the  CCC flux. We expect line contribution
to  be even  smaller in  the case  of  HEG and  LEG, due  to the  wide
spectral  range covered  by the  filter used  ($\sim 2000$  \AA~) with
respect to typical narrow line equivalent widths.  In Table \ref{tab2}
we report  the luminosity  of the  CCCs for the  sample of  FR~II with
$z<0.3$.
  
\begin{table}
\caption{FR~II radiogalaxies nuclei}
\begin{tabular}{l c |l c} \hline
Source Name    &  $\log L_o$       &  Source Name    &  $\log L_o$        \\
               & erg s$^{-1}$ Hz$^{-1}$  &                 &  erg s$^{-1}$ Hz$^{-1}$  \\
               &                &                    &                 \\
   3C~15       &  $<$26.91      &        3C~234      &     29.10       \\
   3C~17       &     28.93      &        3C~236      &  $<$27.04       \\
   3C~18       &     28.62      &        3C~284      &   complex       \\
   3C~33       &    not obs.    &        3C~285      &     25.65       \\
   3C~33.1     &     28.19      &        3C~287.1    &     28.71       \\
   3C~35       &  $<$26.48      &        3C~300      &     27.79       \\
   3C~40       &    complex     &        3C~303      &     28.66       \\
   3C~61.1     &    not obs.    &        3C~318.1    &  $<$25.67       \\
   3C~63       &     28.29      &        3C~319      &  $<$27.40       \\
   3C~79       &     28.33      &        3C~321      &    complex      \\
   3C~88       &     26.60      &        3C~323.1    &     30.14       \\
   3C~98       &  $<$25.67      &        3C~326      &  $<$27.09       \\
   3C~105      &     not obs.   &        3C~327      &  $<$26.55       \\
   3C~111      &     28.14      &        3C~332      &     28.79       \\
   3C~123      &  $<$26.44      &        3C~349      &     28.16       \\
   3C~132      &  $<$26.83      &        3C~353      &  $<$25.53       \\
   3C~133      &     28.00      &        3C~357      &  $<$26.90       \\
   3C~135      &     27.36      &        3C~379.1    &  $<$27.21       \\
   3C~136.1    &    complex     &        3C~381      &  $<$27.52       \\
   3C~153      &  $<$26.66      &        3C~382      &     29.72       \\
   3C~165      &     27.28      &        3C~388      &     27.26       \\
   3C~166      &     28.01      &        3C~390.3    &     29.04       \\
   3C~171      &     26.50      &        3C~401      &     27.62       \\
   3C~173.1    &    complex     &        3C~402      &     26.59       \\
   3C~180      &    complex     &        3C~403      &     26.65       \\
   3C~184.1    &     28.17      &        3C~430      &    complex      \\
   3C~192$^a$  &  $<$26.99      &        3C~436      &    complex      \\
   3C~197.1    &     28.10      &        3C~445      &     29.27       \\
   3C~198      &     28.03      &        3C~452      &  $<$26.89       \\
   3C~219      &     28.64      &        3C~456      &     28.50       \\
   3C~223      &  $<$27.27      &        3C~458      &    not obs      \\
   3C~223.1    &  $<$27.18      &        3C~460      &     27.60       \\
   3C~227      &     28.84      &                    &         	       \\

\hline                                                                        				  
                                                                              				  
\end{tabular}                                                                 
\label{tab2}

\medskip

All observations are  with the F702W filter, except  (a) for which the
F555W  has been  used.  All  luminosities are  calculated at  7000 \AA
~(rest frame) and have  been K--corrected assuming an optical spectral
index $\alpha_o=1$.
                                                                              
\end{table}

\begin{figure}                                 \resizebox{\hsize}{!}
{\includegraphics[width=\textwidth]{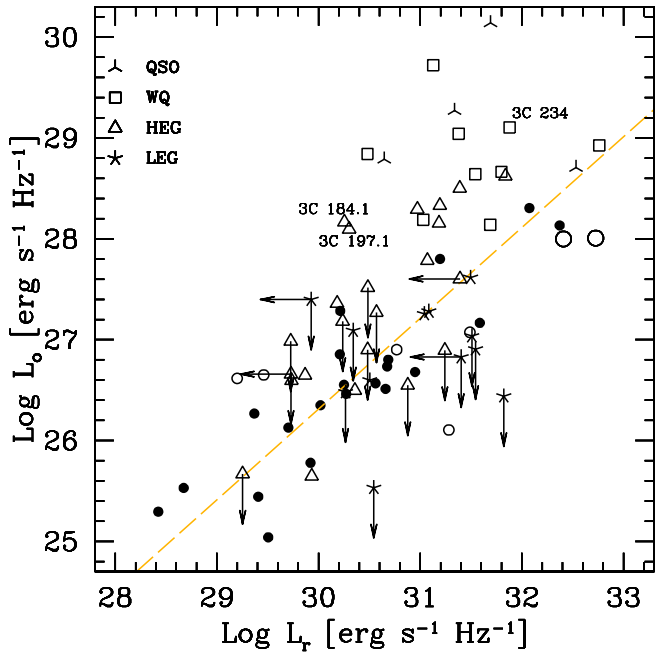}}           \caption{Optical
luminosity of the  CCC vs.  radio core luminosity.   small circles are
FR~I  (open  circles  represent   upper  limits,  filled  circles  are
detections); FR~II are plotted as different symbols depending on their
optical spectral  classification (the  2 large open  circles represent
unclassified objects).  With respect to  Figs.  5 and 3 of paper~I and
II, respectively, we have  removed 3C~386 since from new spectroscopic
observations  of this  object it  has bees  assessed that  the optical
central  source is a  star superimposed  on the  center of  the galaxy
(Marchesini, Celotti \& Ferrarese, in prep.).}
\label{lum}
\end{figure}

Let us now  analyze the behavior of FR~II nuclei  in the plane defined  
by the  radio core luminosity  $L_r$ and optical CCC  luminosity $L_o$  
(Fig.  \ref{lum}),  similarly to what we  have done in  Paper~I and II  
and considering their different optical spectral classification.  
  
The  nuclei of  QSO and  WQ (the  broad-lined objects)  are  among the  
brightest, both in $L_r$ and $L_o$.  They are found exclusively in the  
high  luminosity  region of  the  plane,  for  $L_r >  10^{30.4}$  erg  
s$^{-1}$ Hz$^{-1}$  and $L_o > 10^{28}$ erg  s$^{-1}$ Hz$^{-1}$). Most  
of them  show an  optical excess of  about 2  dex with respect  to the  
radio-optical core  correlation of FR~I CCC.  
Conversely, nuclei of LEGs are only found at low luminosities, i.e. 
for $L_o < 10^{27.5}$ erg s$^{-1}$ Hz$^{-1}$ and $L_r < 10^{32}$ erg 
s$^{-1}$ Hz$^{-1}$ and all lie close to (or just below) the FR~I 
correlation. 
  
The behavior of HEGs is more complex. Their nuclei span a larger range  
of luminosity, both  in radio and optical. Overall,  their CCC (or the  
upper limits) lie  along the FR~I correlation, but  there are at least  
two clear exceptions are  present (namely 3C~184.1 and 3C~197.1) which  
are located well above it.  
  
\section{Discussion}  
\label{discussion} 
  
The behavior of optical nuclei  of the various sub-classes of FR~II in
the radio-optical plane for  this sample essentially confirms what has
been  observed  in  the  lower  redshift sources  discussed  in  Paper
II. However, as the total number of sources here is more than doubled,
we can now  discuss in more detail what the  different position on the
radio-optical  plane  tells  us   about  the  nature  of  the  optical
nuclei. Note  that in the highest  redshift sources, the  CCC might be
contaminated  by a  contribution from  extended features  such  as the
synchrotron  jets  observed  in  some  of the  low  redshift  sources.
However, we are confident that this issue does not significantly alter
our  results,  since these  features  are  observed  only in  a  small
fraction of FR~I  galaxies (possibly only in those  with jets pointing
towards  us).  Furthermore,  the contribution  from the  observed jets
(e.g. in 3C~264) is typically less than 20\% of the CCC flux.

\subsection{QSO and WQ: the broad-lined objects}  
  
Let  us  firstly examine  the  case of  WQ  and  QSO, the  broad-lined  
objects. As discussed  in Paper~II, the optical excess  shown by these  
sources, with respect to the  radio core emission, can be explained if  
the dominant component in the  optical band is due to thermal emission  
from an accretion disc.  
  
  
A very  important result here  is that the  optical nuclei of  BLO are
present  only for  $L_{\rm o}  > L_{{\rm  o, min}}  \sim  10^{28}$ erg
s$^{-1}$ Hz$^{-1}$,  corresponding to  a luminosity $\nu  L_{\nu} \gta
4\times 10^{42}$ erg s$^{-1}$.  A rather different behavior is seen in
radio-quiet AGN,  in which broad lines  are seen in  objects that span
many orders of magnitude in nuclear optical luminosity, from LINERS to
powerful QSO.  Recent  works have indeed shown that  in LINERS (and in
several  Seyfert~1  galaxies)  the  nuclear bolometric  luminosity  of
Type~1 radio quiet  AGNs can be as low as  $L_{bol} \sim 10^{40}$ (see
e.g.   Barth   et  al.   \cite{barth}  for  NGC~4579,   Moran  et  al.
\cite{moran} for NGC~4395).

We are  quite confident that the lack  of BLO below this  limit is not
due  to selection  effects.  In  fact, although  spectral data  of 3CR
radio galaxies are not homogeneous, Willott et al. \cite{willott} have
shown that  in these low  redshift sources the most  prominent optical
broad line (H$\alpha$) should be easily detected if the observed trend
between the extended  jet and emission line power  extends smoothly to
the lower luminosity  objects (Rawlings \& Saunders \cite{rawlings91},
Celotti et al.  \cite{celpadgg97}).
  
If  the optical nucleus  observed in  BLO is  directly related  to the
inner disk, we  can derive some important parameters  of the accretion
process, under the  assumption that the optical luminosity  is a fixed
fraction of $L_{\rm bol}$, the  bolometric luminosity of the AGN. This
translates  to   $L  \gta  3  \times  10^{-4}   \eta  L_{Edd}$,  where
$\eta=L_{\rm bol}/L_o$ and $L_{\rm Edd}= 1.3 \times 10^{46} M_{8}$ erg
s$^{-1}$  is the  Eddington luminosity  for a  $10^8  M_{\odot}$ black
hole.   For any fixed  value of  the radiative  efficiency $\epsilon$,
this in turn  translates in an accretion rate $\dot m  \equiv \dot M /
\dot  M_{Edd} \simeq  3 \times  10^{-4} M^{-1}_8  \epsilon^{-1} \eta$.
Assuming $\epsilon  = 0.1$, i.e. radiatively  efficient accretion, and
$\eta\sim 15$,  following Elvis et  al.  (\cite{elvis}), the  value of
the  accretion rate  corresponds to  $\dot m  \simeq 5  \times 10^{-2}
M^{-1}_8 \epsilon_{0.1}^{-1} \eta_{15}$.  Note, however, that Zheng et
al. (\cite{zheng1997}) have  recently found that the peak  of the disk
radiated  power  occurs at  a  much  lower  frequency than  previously
evaluated, and  possibly only  a factor of  2 higher than  the optical
luminosity.  This implies,  for a  fixed optical  luminosity,  a lower
value of $L_{\rm bol}$ and in turn a lower accretion rate.
   
A possible scenario  for the presence of a lower  limit in the optical 
luminosity  of BLO  is  that our  estimate  of $\dot  m$ represents  a 
threshold  at which  the accretion  process changes  its  regime.  For 
lower values of $\dot m$ this accretion might occur at a low radiative
efficiency, in the form of an advection dominated flow (ADAF, see e.g.
Narayan \&  Yi \cite{narayan95}), an  adiabatic inflow-outflow (ADIOS,
Blandford \& Begelman  \cite{bland99}), or a convection-dominated flow
(CDAF, Narayan,  Igumenshchev, \& Abramowicz  \cite{cdaf00}). As shown
in Paper I, FR~I radio-galaxies,  excluding the small minority of them
that  show  clear  sign  of  quasar activity  (e.g.   3C~120),  might
represent examples of such low efficiency processes.  The lower amount
of ionizing photons available would  account for the lack (or at least
for the substantial lower intensity) of broad lines in these objects.
  
However, the presence of an accretion threshold generates a 
corresponding threshold in luminosity only if there is a relatively 
narrow range of black hole masses at a fixed radiative efficiency. 
While a wide range of black hole masses are likely to be harbored in 
radio-quiet AGNs, we then speculate that black hole masses in 
radio-loud AGN could be only restricted to a relatively small range, 
at the higher end of the black hole mass distribution.  This seems to 
be confirmed observationally, as the measured masses in these objects 
are all concentrated around 10$^9 M_\odot$ (e.g.  Dunlop et 
al. \cite{dunlop}, Laor \cite{laor}).

\subsection{High Excitation Galaxies}  
  
  
In the standard unification  picture, narrow--lined FR~II are believed  
to  be   intrinsically  identical   to  the  broad--lined   ones,  and  
obscuration of the central regions is assumed to be the only origin of  
the difference  in the  emission line properties  of the  two classes.  
Note that this is strictly true only in the case of HEG, for which the  
narrow line  properties (luminosity and excitation  state) are similar  
to those of WQ (and QSO).  In other words, HEGs are expected to harbor  
a hidden quasar, while for LEGs this is not strictly required.  
  
The nature of the nuclei of HEG is puzzling. Since most of them lie on
the FR~I correlation, they might be interpreted, in analogy with FR~I,
as due to  non thermal synchrotron emission from the  base of the jet.
Of course this  appears to be in contrast  to the standard unification
scenario.  However,  two narrow-lined objects  (3C~184.1 and 3C~197.1)
clearly stand out,  showing an optical excess of more  than 1 dex when
compared  to sources  with similar  radio core  power.  The  nature of
these objects  is unclear: in  3C~184.1 a broad (although  very faint)
Pa$\alpha$  line  has been  detected  while  no  broad components  are
observed in  optical lines (Hill  et al.  \cite{hill96}).   A possible
explanation is that they are moderately absorbed QSO nuclei.
  
Another  puzzling source  is 3C  234.  Spectropolarimetric  studies of
this source showed that its polarized spectrum closely matches that of
a QSO, with prominent featureless continuum and broad lines (Antonucci
\cite{antonucci90},   Tran   et    al.    \cite{tran1995},   Hurt   et
al.  \cite{hurt}).   This  has  been  interpreted  as  the  result  of
scattering of light from a hidden  QSO nucleus into our line of sight.
Nonetheless, its representative point  lies on the correlation (within
2$\sigma$).   We  have calculated  that  the  scattered light  exactly
corresponds to the flux measured from the nuclear component in the HST
images  of   3C~234\footnote{The  polarization  level   of  its  broad
H$\alpha$  line is  $\sim 20$  \%, after  removal of  the  host galaxy
starlight.   This  indicates that  the  total  flux  of the  scattered
H$\alpha$  is   approximately  5  times  larger   than  its  polarized
component.  Applying the same  correction to the featureless continuum
emission  level we  infer  a  scattered flux  of  $\sim 10^{-16}$  erg
s$^{-1}$ cm$^{-2}$ A$^{-1}$.  The CCC flux in the  HST image of 3C~234
is $1.1  \times 10^{-16}$ erg  s$^{-1}$ cm$^{-2}$ A$^{-1}$.}.   We can
then argue that  the nuclear source in 3C 234  is a compact scattering
region  (which   appears  as   unresolved  also  in   high  resolution
polarimetric   images  obtained   with  the   Keck   telescope,  Cohen
\cite{cohen99}).  The  same argument applies to 3C~109,  a FR~II radio
galaxy with a  redshift of z=0.3056, just above  the distance limit of
our sample.   Its polarization properties are reminiscent  of those of
3C 234 and  it was similarly interpreted as a  hidden quasar.  Also in
this case, we have tested that the scattered continuum matches its CCC
flux\footnote{The  continuum flux  in 3C~109  is as  polarized  as the
broad  lines  (indicating  the   absence  of  a  significant  diluting
component) at a level of $\sim  7$ \%.  Given a polarized flux at 7000
\AA\ of $\sim 2 \times  10^{-17}$ erg s$^{-1}$ cm$^{-2}$ A$^{-1}$ this
translates into a scattered component  of $\sim 3 \times 10^{-16}$ erg
s$^{-1}$ cm$^{-2}$ A$^{-1}$. The nuclear  source in the HST images has
a  flux of $2.8  \times 10^{-16}$  erg s$^{-1}$  cm$^{-2}$ A$^{-1}$.}.
Remarkably,  its  CCC  also  lies  on the  correlation  (again  within
2$\sigma$).
  
Therefore,  besides  synchrotron   radiation  from  the  jet,  another 
possible scenario for  the nature of such nuclei  is that the observed 
optical emission might originate from the expected (obscured) quasar 
component,  which is  scattered to  our line  of sight  by ``mirrors'' 
placed out of the obscuring torus. 
  
From  this  analysis  it  turns  out  that it  is  very  difficult  to
definitively  address the  nature of  HEG nuclei  by  only considering
their position in the optical--radio plane, particularly in the region
of high  radio core luminosity, where the  FR~I correlation intersects
the BLO region.  In order to clarify their nature, more information is
needed.   As   we  show  in  the  following   section,  a  fundamental
advancement will be  achieved with the inclusion in  our analysis of a
further parameter, i.e.  the luminosity of their emission lines.

\subsection{EW of [OIII] vs. the radio optical correlation:   
a new diagnostic plane for radio galaxies}  
\label{fplane}

\begin{figure}   
\resizebox{\hsize}{!}{\includegraphics[width=\textwidth]{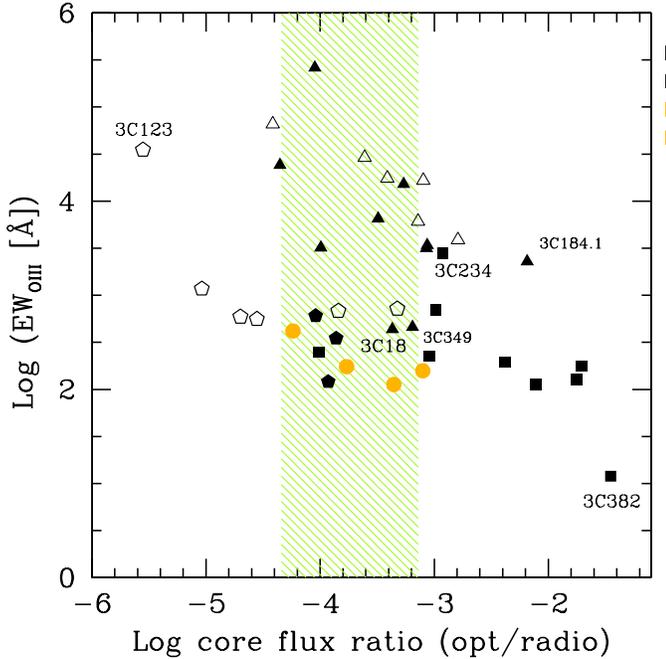}}   
  
\caption{Equivalent width  of the [OIII] emission  line, measured with  
respect to  the CCC  emission, is plotted  vs.  the ratio  between the  
optical  CCC  to  radio  core  flux.   Filled  symbols  represent  CCC  
detections, while  empty symbols are  upper limits.  Squares  are BLO,  
triangles are HEG, pentagons are  LEG and circles are FR~I. The shaded  
area represents  the dispersion (1$\sigma$) of  the linear correlation  
between radio core and CCC luminosity found for FR~I.}  
  
\label{ew}  
\end{figure}

In Fig.  \ref{ew} we plot the ratio between the [O III] flux and the 
optical flux of the CCC, which essentially represents the equivalent 
width of the emission line with respect to the nuclear component, 
versus the radio core -- CCC ratio. This emission line appears in fact 
to be strictly connected to the nuclear ionizing source.  The value of 
the [OIII] flux is taken by Jackson \& Rawlings \cite{jackson} and 
from the data collected by Chris Willott in his home page 
(http://www-astro.physics.ox.ac.uk/~cjw/3crr/3crr.html).  We have 
collected [O III] line fluxes for 42 FRII and for 5 FR~I, while 
complete information (line flux, radio and optical core, either 
detections or upper limits) is available for 33 FR~II and 4 FR~I.  The 
location with respect to the X axis is readily understood noting that 
the radio--optical core correlation of FR~I is represented by a fixed 
radio-to-optical ratio, i.e. a vertical stripe in this plane, centered 
on $F_o/F_r \sim 10^{-3.7}$.  The 1$\sigma$ region around this value 
is represented in the figure as the shaded area. Objects lying on the 
right side of the plot show an optical excess. 
  
In this plane, sources separate in two regions corresponding to 
equivalent width values EW $\sim 10^{2.5}$ \AA~ or EW $>10^{3.5}$ \AA. 
In the first region we find all QSO and WQ (squares) 
\footnote{The  only BLO  with value  of EW  higher than  1000  \AA ~is  
3C~234 which, as already pointed out in section \ref{sample},  
should be more properly re-classified as a HEG.}  and LEG (pentagons).  
Interestingly, also the four FR~I (3C~66B, 3C~84,  
3C~346 and  3C~449, plotted  as filled grey  circles) for  which these  
data are available  fall into this region. Conversely,  all but two of  
the HEG  have much larger  equivalent widths, from EW  $\sim 10^{3.5}$  
\AA~ to EW $\sim 10^{5.5}$ \AA.

\begin{figure}  
   \resizebox{\hsize}{!}  
 {\includegraphics[width=\textwidth]{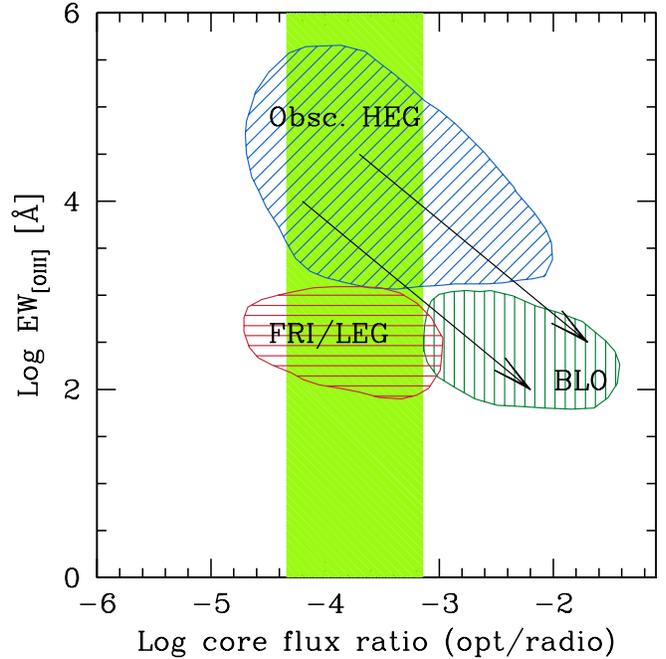}}  
\caption{Different classes of radio galaxies occupy different regions 
in  their ``fundamental'' diagnostic plane,  made by  the EW  of the  [OIII] line 
plotted vs.  the optical  to radio core  flux ratio.  ``Obscured'' HEG 
have the highest EW. These sources can be lead back to the region of BLO  
if their optical flux is increased by a factor of $\sim 100$ (arrows).}  
\label{fagioli}  
\end{figure}

The separation of different  sources depending on the equivalent width
of their  emission line is  readily expected from the  unified models.
This is  due to the effects  of obscuration that  strongly reduces the
observed  nuclear  continuum, while  the  line  emission, produced  at
larger distance  from the nucleus,  is unaffected.  We can  then argue
that all  sources with EW  $> 1000$ \AA~ are  hidden BLO\footnote{Note
that since the  [OIII] emission line regions have  a typical dimension
of  10 kpc  (e.g.  Baum \&  Heckman  \cite{baumheckman89}), such  high
values  of the EW  are still  compatible with  the assumption  made in
Sect.   \ref{fr2ccc} of  a small  line contamination  for  the nuclear
flux.}.  In sources with very high values of EW (EW $> 10^{3}$ \AA), a
strong  ionization source,  obscured  to our  viewing  angle, must  be
present.  This argument follows essentially the same line of reasoning
that has  been previously  applied to Seyfert  galaxies for  which, in
several cases,  a deficit  of ionizing photons  with respect  to their
line luminosity  has been found  (e.g. Kinney et  al.  \cite{kinney}).
Assuming the median value of the  EW $\sim 10^{2.5}$ \AA~ for BLO, the
spread in the  EW of HEG corresponds to  a scattering fraction varying
between 0.001  and 0.03.  Indeed note that  obscuration, which affects
only the  optical luminosity, moves  the objects along  straight lines
parallel to the diagonal of the plane.  Indeed, all galaxies with high
values of EW can be led  back to the region typical of QSO increasing
their nuclear luminosity  by, on average, a factor  of $\sim$ 100 (see
Fig. \ref{fagioli}).
 
The sources with  low values of EW are all  clustered around EW $\sim$
300 \AA.  This value is essentially independent of the source class of
luminosity. A low value of EW thus appears to be the signature that we
are  directly  observing the  source  of  ionization.   While this  is
straightforward for  QSO and  WQ, it implies  that in FR~I  either the
source of gas ionization is  the synchrotron emission from the jet, or
any other  contribution must have  an intensity comparable to  the jet
emission. Since this must hold  for all radio luminosities, the latter
hypothesis appears to be quite  finely tuned. The correlation found by
Verdoes Kleijn  et al.  (\cite{kleijn}) between  the nuclear H$\alpha$
luminosity and the optical CCC luminosity in FR~I further supports the
former scenario.
  
Only  two HEG  lie  among the  ``unobscured'' sources:  interestingly,
these galaxies (3C~18 and 3C~349) also lie on the FR~I correlation, on
its  high luminosity  end. Therefore,  due to  their position  in both
planes, we can  interpret that these galaxies are  true unobscured and
synchrotron dominated ``FR~I--like'' FR~II.
 
Since this  plane allow us to  separate and to identify  the nature of 
the nuclear sources of both FR~I  and FR~II, we consider this as a new 
``fundamental'' diagnostic  plane  for   radio  galaxies.     
 
Another interesting  consequence of  this new characterization  is the 
possibility  of inferring  the  nature  of the  sources  in which  the 
nucleus  is  not  detected.   Upper  limits among  the  HEG  would  be 
attributed to obscuration of the central quasar--like source and small 
amount  of  scattered  radiation  (compared  to  the  central  surface 
brightness of  the host galaxy).   In the case  of LEG, this  issue is 
more complex: as  we expect their nuclei to  be fainter, the ambiguity 
between low contrast and even mild absorption holds.  However, in case 
of absorption  by a torus--like  structure in LEG, objects  with upper 
limits   and   detections   should   have  a   different   orientation 
distribution. In order to address this issue, and before proceeding to 
discuss  the properties  of  LEG  and their  role  in the  unification 
models, it is necessary to analyze how the different classes relate to 
the radio properties.

\subsection{Relationship with the radio properties}  
\label{orientation}

In  this  section  we   compare  the  distribution  of  the  different
subclasses of FR~II  both in total radio luminosity at  178 MHz and in
$R$, the ratio between core and total radio power, which is often used
as an orientation  indicator.  This is crucial both  to understand the
role of the  different sources in the unification  scheme, and to know
whether orientation plays  a role in the presence  (or absence) of the
CCC.
  
\begin{figure}  
   \resizebox{\hsize}{!}  
   {\includegraphics[width=\textwidth]{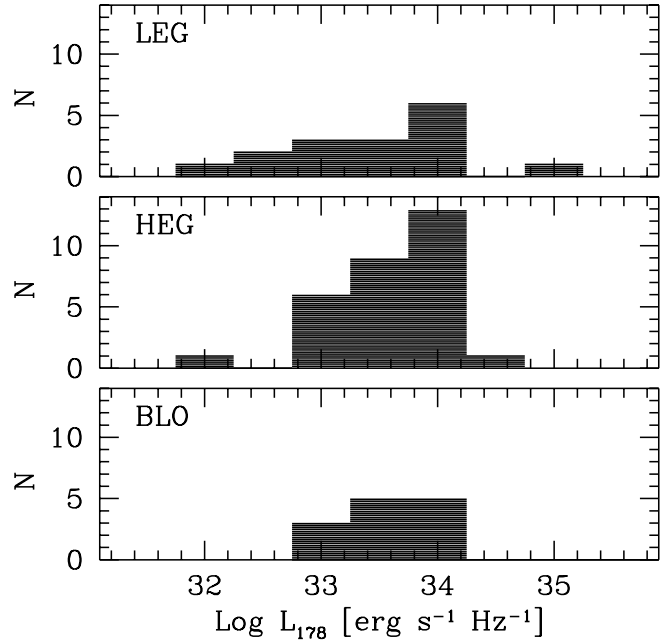}}  
   \caption{Histograms of the radio total luminosity at 178 MHz for the   
three spectral classes of FR~II.}  
              \label{istolext}  
    \end{figure}

Although a similar analysis has been already done on a wider sample of
3CR sources by  Laing at al.  \cite{laing94}, it  is very important to
perform statistical tests on our  sample, taking also into account the
newly obtained  information on the  optical nuclear properties  of our
objects.  In  order to improve  the statistics, we have  considered WQ
and QSO as a single  class, due to their similar properties.  However,
we have checked that this does  not affect the results, as the objects
belonging  to these  two classes  also share  the same  range  in both
$L_{178}$ and  $R$.  In Fig.  \ref{istolext} we  show the distribution
in $L_{178}$  for the three different spectral  classes.  Although the
number of objects for each  class is limited, it is straightforward to
note that: i)  they share the same range in total  power; ii) they are
similarly  distributed. In  particular, this  implies that  within our
range of redshift the three subclasses of FR~II co-exist and cannot be
discriminated on the basis of their total radio luminosity.

   \begin{figure}  
   \resizebox{\hsize}{!}  
   {\includegraphics[width=\textwidth]{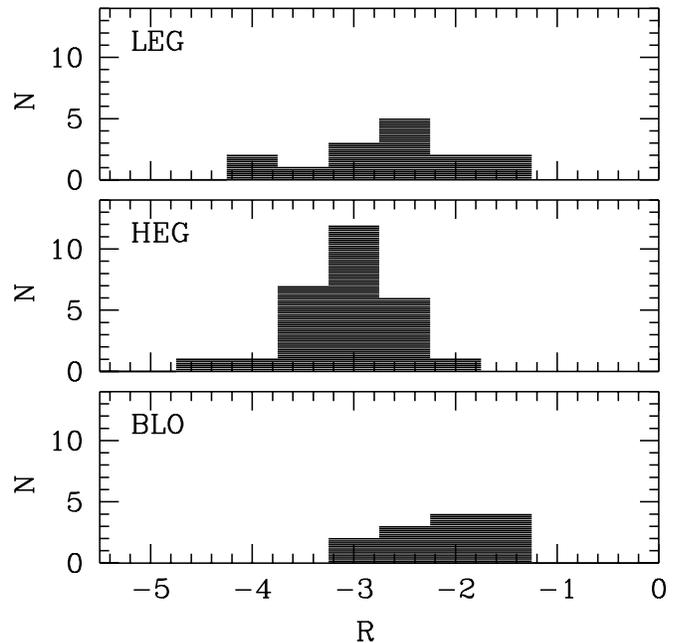}}  
   \caption{Histograms of the radio core (at 5 GHz) to total flux (at 178 MHz)  
for the three spectral classes of FR~II.}  
              \label{istor}  
    \end{figure}

We use a Kolmogorov--Smirnov test  to see whether the distributions of 
$R$ for the various sub-classes, as well as for objects with different 
nuclear properties, are statistically different. 
From this analysis, we deduce  the following:  
  
$\bullet$ broad-lined objects have higher $R$ with respect to both HEG  
($P > 99.96\%$) and LEG ($P > 88\%$).  This fits in the framework in which  
BLO are sources  where the nuclear emission is  seen directly, and the  
observing angle is smaller than the  HEG one (but still larger than in  
beamed blazar sources).  Their high luminosity CCCs can be produced in  
the very inner regions of the AGN, possibly in the accretion disc;  
  
$\bullet$ the distributions in $R$ of  HEG and LEG are different ($P >
91.9  \%$).  This  is  consistent  with the  conclusions  of Laing  et
al. (\cite{laing94}) that LEG constitute a randomly oriented sample;
 
$\bullet$  An important  result  here is  that  HEG and  LEG with  and 
without CCC  do not have  different distributions in $R$.   This might 
imply  that the  presence  of CCC  in  these sources  does not  depend 
strongly on orientation.  Note that this is quite  obvious for HEG, if 
the origin of the majority of their nuclei is scattered radiation. 
 
In  light of  these results,  we  have acquired  enough information  to 
discuss the properties of LEG. 
 
\subsection{The Low Excitation Galaxies}  
  
As already pointed out above, the nuclei of LEG, when detected, lie on 
the FR~I  correlation, and they  also have EW  very close to  those of 
FR~I. From the  above statistical analysis, we also  confirm that LEGs 
in our sample  have a broad distribution in  $R$, compatible with that 
of a randomly oriented population.   Therefore, from the point of view 
of   their  radio-optical  nuclear   properties,  these   objects  are 
indistinguishable  from  FR~I.   Since   their  nuclei  appear  to  be 
dominated by synchrotron radiation from the relativistic jet, the lack 
of strong (and high ionization)  emission lines in their spectra might 
be  accounted  for by  a  lower  number  of ionizing  photons.    
 
These objects, when  observed along the jet axis,  should appear as BL
Lac objects and can therefore account for the known BL Lacs with FR~II
radio morphology.  Note  that this is in contrast  with a ``standard''
unification model,  which associates all  FR~II to quasars,  but these
objects  can well  account for  BL Lacs  with both  an  extended radio
morphology  (e.g. Kollgaard et  al.  \cite{kollgaard92})  and extended
radio power (Cassaro et al. \cite{cassaro}) typical of FR~II.  Jackson
\& Wall  \cite{jacksonwall99} have also shown that  both the evolution
and  unification of  FR~I and  FR~II with  BL Lacs  and flat--spectrum
quasars (FSRQ)  are well accounted  for by a  dual--population scheme,
which considers FR~I  and LEG as a single  population, associated with
BL Lacs, while all other FR~II  are unified to FSRQ. Thus, our results
are in agreement with such a scenario.
  
Some independent  pieces of evidence  appear to confirm  this picture. 
Recently, Willott et al. (\cite{willott}) have found that the fraction 
of quasars  in the  3CRR, 6C and  7C samples (having  excluded objects 
with FR~I  radio morphology) decreases  as luminosity (both  radio and 
narrow line) decreases; this behavior  might be explained in the framework 
of  a ``two  population'' model:  at low  luminosities,  a significant 
fraction  of  sources  are  ``starved  quasars'',  and  their  nuclear 
properties,  as well  as  the characteristics  of  their spectra,  are 
similar to those  of FR~I.  It is therefore  tempting to identify this 
class with LEGs. 
 
We must point out that a particular source (3C~123) does not appear to 
fit  this scenario, since  it has  no optical  CCC and  its equivalent 
width of [OIII] is higher  than $10^3$ (see Fig.  \ref{ew}).  However, 
its EW is  larger by more than one order of  magnitude compared to all 
of the  other LEG, and also  its O[III] luminosity is  more typical of 
HEG.  Unfortunately,  no detailed spectral information  is available in 
the  literature  for this  object.   In  particular, the  [OIII]/[OII] 
ratio, which is a  fundamental parameter for a correct classification, 
is unknown.  Therefore, in light  of its position in the diagnostic 
plane, we argue that that 3C~123 should be re-classified as a HEG.

\section{Summary and conclusions}  
\label{conclusions} 
  
We have analyzed  the optical nuclear properties of  a complete sample
of 65  FR~II radio galaxies up  to $z=0.3$ from the  3CR catalog.  The
overall  scenario  basically  confirms  the  findings  for  the  lower
redshift   sample   ($z<0.1$)    presented   in   Chiaberge   et   al.
(\cite{pap2}).   However,  the larger  number  of  sources (more  than
double) allows us to reveal  a richer and more complex behavior, which
turns  out  to  be   closely  associated  with  the  optical  spectral
properties of the different objects.  The nuclear properties of FR~II,
as inferred from our analysis, can be summarized as follows.
  
While the  great majority of  FR~I radio galaxies (optical  and radio)
nuclei  are dominated  by  non-thermal synchrotron  emission from  the
relativistic jet,  the FR~II population is  not homogeneous.  Although
optical Central  Compact Cores appear to  be a common  feature also in
FR~II  galaxies, their origin  can be  ascribed to  different physical
processes.
  
BLO typically  have the  brightest nuclei which  show a  large optical
excess with  respect to the radio-optical correlation  found for FR~I.
This is  readily explained  if the dominant  component in  the optical
band is due to thermal emission from an accretion disc.  We found that
optical nuclei of BLO are present only for $L_{{\rm o}} > 10^{28}$ erg
s$^{-1}$  Hz$^{-1}$.    A  rather  different  behavior   is  seen  in
radio-quiet AGN,  in which broad lines  are seen in  objects that span
many orders of magnitude in nuclear optical luminosity, from LINERS to
powerful  QSO. As we  argued that  this effect  cannot be  ascribed to
obscuration  or  selection  effects,  we  suggest  that  this  is  the
manifestation  of  a threshold  in  the  efficiency  of the  accretion
process,  from  the   standard  optically  thick,  geometrically  thin
accretion disk to low radiative accretion flows.
  
Note that the presence of any limit in luminosity requires a well 
defined behavior of the accretion rate and radiative efficiency but 
also a narrow distribution in black hole masses for radio-loud AGNs. 
This conjecture seems indeed to be supported by the direct 
measurements available to date. 
  
The nature of  the nuclei of HEG is  certainly more complex.  Although
most sources  lie along the FR~I  correlation, there are  at least two
clear exceptions, showing  a significant optical excess.  Furthermore,
there  are  at  least  two  sources  (3C~234  and  3C~109)  for  which
spectropolarimetric studies  clearly showed that they  harbor a hidden
QSO nucleus.  The amount  of scattered continuum light matches exactly
the flux of their nuclear component in the HST images, indicating that
their  CCC  is  a   compact  scattering  region.   Nonetheless,  their
representative  points would  lie on  the correlation  by coincidence.
From this  analysis it turns out  that in general it  is impossible to
definitively  address the  nature of  HEG nuclei  by  only considering
their position in the  optical--radio plane.  A fundamental advance is
achieved with  the inclusion in  our analysis of a  further parameter,
i.e. the luminosity of their emission lines.
  
In the plane formed by the ``nuclear'' EW of the [OIII] line vs the 
optical excess with respect to the non-thermal jet emission (a new 
``fundamental'' diagnostic  plane?), the different classes of sources clearly 
separate according to the nature of their nuclei.  For low EW values 
($\sim 10^{2.5}$ \AA) we find all QSO, WQ, LEG and FR~I, which differ 
only by the amount of the optical excess.  On the other hand, all but 
two of the HEG have much larger equivalent widths \gta $10^{3.5}$ \AA. 
The separation of sources depending on their line equivalent width is 
indeed expected from the unified models, as obscuration reduces the 
nuclear continuum emission while the line emission is less affected or 
unaffected.  In sources with very high values of EW a strong 
ionization source, obscured to our viewing angle, must be present.  We 
can then argue that all sources with high EW are hidden BLO. 
  
The low EW region would then contains objects in which we see directly  
the source  of ionization.  A ramification  of this result  is that in  
FR~I the most likely dominant  source of gas ionization is synchrotron  
emission from the jet.  
  
Only   two  HEG   are  located   among  the   ``unobscured''  sources:  
interestingly, these galaxies (3C~18 and  3C~349) also lie on the FR~I  
correlation.   Therefore, due  to their  position in  both  planes, we  
identify these objects with true unobscured ``FR~I--like'' FR~II.  
  
According to the scenario proposed  here, the non--detection of CCC in  
galaxies of  different classes should have different  origins. For LEG  
this  might be  due to  either a  low contrast  with the  stellar host  
galaxy  emission or  to a  (moderate) amount  of  absorption, randomly  
oriented with respect to the jet.  On the other hand, for HEG this can  
be only attributed to a lower amount of scattered nuclear radiation.  
  
The  picture which  emerges is  that  radio galaxies  manifest in  two
types,  which are  not  directly related  to  the extended  FR~I/FR~II
dicothomy.   In  the framework  of  the  unification  scheme, BLO  and
obscured  HEG  appear to  have  the  same  nuclear structure:  intense
thermal  disk  (ionizing) emission,  substantial  broad emission  line
region,  torus-like absorber and,  of course,  powerful jets.   On the
other  hand,  LEG,  FR~I  and  unabsorbed HEG  constitute  a  distinct
population, characterized  by low radiative  efficient accretion, weak
or absent broad line emission, lack of a significant nuclear absorbing
structure. Unfortunately we  have complete information (radio, optical
and emission line) for only roughly half of the sources in our sample.
Although we can  {\it only} make the assumption that  there is no bias
in their  selection, we can estimate  that the population  of FR~II is
composed of  $\sim 50\%$ obscured sources harboring  a quasar nucleus,
$\sim  25\%$ BLO,  $\sim  20\%$ LEG  and  $\sim 5  \%$ FR~I--like  HEG
(although so far  this is assessed only for  two sources).  The latter
two classes can account for BL Lac objects with FR~II radio morphology
and extended  radio power (e.g. Kollgaard  et al.  \cite{kollgaard92},
Cassaro  et  al.   \cite{cassaro}).   This scenario  apparently  poses
problems  for  the  simplest  unification  models,  in  particular  in
identifying  the  beamed  counterparts  of FR~I--like  HEG.   However,
strong and  high excitation narrow emission lines  are indeed observed
in  few BL  Lacs (Landt  et  al.  \cite{landt}).   Moreover, the  line
equivalent width  in the  beamed objects will  be reduced by  a factor
$\sim 10^4$ (the typical  ratio between the nuclear optical luminosity
between   radio   galaxies  and   BL   Lacs,   see   Capetti  et   al.
\cite{capetti02})   producing  values   consistent  with   a   BL  Lac
classification.

A possible way to test the  overall picture is to look at the spectral
properties of the nuclei. BLO  and scattered nuclei are expected to be
different  from   the  FR~I   and  FR~I--like  synchrotron   ones:  in
particular, we expect to  observe flatter spectral indices, typical of
quasars, in BLO, indicating the  presence of a thermal blue bump.  Due
to  the  large  uncertainties,   optical  observations  (even  in  two
different bands) are not enough to determine the spectral slope, which
instead  could  be better  measured  by  taking  advantage of  the  UV
information (Chiaberge et al. \cite{papuv}).  In addition, an infrared
nuclear excess is expected in  the obscured radio galaxies, while this
has  to be  absent  in  FR~I--like objects,  as  promisingly shown  by
Whysong \& Antonucci  (\cite{whysong}) for the case of  3C~405 (a true
obscured quasar)  and 3C~274 (M~87).  A further diagnostic tool  is of
course  the observations  in the  X--ray  band, which  can reveal  the
presence of different amounts of nuclear absorption.
   
Finally,  it would  be  particularly interesting  to  analyze how  the
properties  of  the  newly   discovered  quasars  showing  FR~I  radio
morphology  (Blundell   \&  Rawlings  \cite{blundraw},   Lara  et  al.
\cite{lara}) fit in  our picture. In a sense  they might represent the
analogous (but  oposite) case of LEG,  where an FR~II  harbors an FR~I
nucleus.   The spectral  information available  in the  literature for
such objects  is, to our best  knowledge, not yet  sufficient to state
where  these   sources  are  located   in  the  diagnostic   plane  of
Fig.~\ref{ew}. However, if they  are indeed broad--lined FR~I (showing
optical  thermal emission  from  the accretion  disk)  they should  be
placed in the  lower--right end of the plane, among  the BLO.  If this
is  the case,  the properties  of these  peculiar sources  might bring
further support to the models that claim that the nuclear structure is
not  directly  connected  to   the  extended  radio  morphology  (e.g.
Bicknell   \cite{bicknell84,bicknell94};    Gopal-Krishna   \&   Wiita
\cite{wiita}).

\begin{acknowledgements}

The authors thanks F.~D.   Macchetto for insightful discussions.  This
research has  made use of  the NASA/IPAC Extragalactic  Database (NED)
which  is  operated  by  the  Jet  Propulsion  Laboratory,  California
Institute of Technology, under  contract with the National Aeronautics
and Space Administration.

M.C. wish  to thank  S.  Crawshaw,  B. Waghorn and  A. De  Martino for
stimulating conversations.

\end{acknowledgements}

\end{document}